# The Cryogenic Anticoincidence Detector for ATHENA X-IFU: Preliminary test of AC-S9 towards the Demonstration Model

M. D'Andrea[*,a,b], C. Macculi[a], A. Argan[a], S. Lotti[a], G. Minervini[a], L. Piro[a],
M. Biasotti[c], V. Ceriale[c], G. Gallucci[c] F. Gatti[c], G. Torrioli[d], A. Volpe[e]

[a]INAF/IAPS Roma, Via del Fosso del Cavaliere 100, 00133 Roma (Italy);
[b]Dept. of Phys. Univ. of Rome "Tor Vergata", Via della Ricerca Scientifica 1, 00133 Roma (Italy);
[c]Dept. of Phys. Univ. of Genoa, Via Dodecaneso 33, 16146 Genova (Italy);
[d]CNR/IFN Roma, Via Cineto Romano 42, 00156 Roma (Italy);
[e]ASI, Via del Politecnico, 00133 Roma (Italy);

## ABSTRACT

Our team is developing the Cryogenic Anticoincidence Detector (CryoAC) of the ATHENA X-ray Integral Field Unit (X-IFU). It is a 4-pixels TES-based detector, which will be placed less than 1 mm below the main TES array detector. We are now producing the CryoAC Demonstration Model (DM): a single pixel prototype able to probe the detector critical technologies, i.e. the operation at 50 mK thermal bath, the threshold energy at 20 keV and the reproducibility of the thermal conductance between the suspended absorber and the thermal bath. This detector will be integrated and tested in our cryogenic setup at INAF/IAPS, and then delivered to SRON for the integration in the X-IFU Focal Plane Assemby (FPA) DM.

In this paper we report the status of the CryoAC DM development, showing the main result obtained with the last developed prototype, namely AC-S9. This is a DM-like sample, which we have preliminary integrated and tested before performing the final etching process to suspend the silicon absorber. The results are promising for the DM, since despite the limitations due to the absence of the final etching (high thermal capacity, high thermal conductance, partial TES surface coverage), we have been able to operate the detector with $T_B$ = 50 mK and to detect 6 keV photons, thus having a low energy threshold fully compatible with our requirement (20 keV).

## 1. INTRODUCTION

The Advanced Telescope for High ENergy Astrophysics (ATHENA [1]) is the second large-class mission selected in ESA Cosmic Vision 2015-2025, with a launch foreseen in 2031 towards an L2 halo orbit. One of the two focal-plane instruments is the X-ray Integral Field Unit (X-IFU [2]), a cryogenic spectrometer able to simultaneously perform high-resolution spectroscopy ($\Delta E$ < 2.5 eV at 7 keV) and detailed imaging (5" angular resolution with a 5' diameter FoV) over the 0.2-12 keV energy band. It is based on a large array (about 4000 pixels) of Transition Edge Sensor (TES) microcalorimeters, working at a bath temperature of 50 mK. To reduce the particle-induced background, thus enabling part of the mission science goals, the instrument includes a Cryogenic Anticoincidence detector (CryoAC) [3]. It is a 4-pixels TES-based detector, placed less than 1 mm below the TES array (Fig. 1).

The X-IFU development program foresees to build and characterize an instrument Demonstration Model (DM) before the mission adoption. In this respect, we are now developing the CryoAC DM, which will be delivered to the Focal Plane Assembly (FPA) development team at SRON for the integration with the TES array DM. This will be the first compatibility test for the two detectors, representing a milestone on the path towards the X-IFU development.

* mail: matteo.dandrea@iaps.inaf.it

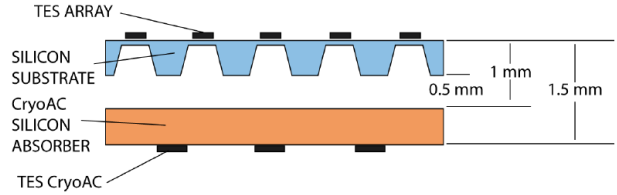

Figure 1. X-IFU TES array to CryoAC distance sketch.

At subsystem level, the CryoAC DM is part of a wider program aimed to reach TRL 5. This will be performed developing two representative detector models:

- The Demonstration Model (DM): a suspended square single pixel able to probe the detector critical technologies;
- The Structural Thermal Model (STM): a mechanical model of the CryoAC 4-pixels array to be vibrated and cooled in a representative environment.

Here we will first present the CryoAC DM, describing the detector design and the fabrication process. Then we will report the main results obtained with AC-S9, a DM-like prototype that we have integrated and tested before performing the final etching process to suspend the silicon absorber. This has allowed us to verify all the previous fabrication steps, and it will allow us to check if the final etching can affect the detector.

## 2. THE CRYOAC DEMONSTRATION MODEL

The main requirements to be satisfied in the DM context to enable the CryoAC critical technology are:

- Pixel size (absorber area) = $10 \times 10$ mm$^2$, to be representative of the Flight Model (FM) pixel dimensions (4 pixels, ~1.2 cm$^2$ area each one);
- Low energy threshold ≤ 20 keV, necessary to meet the requirement on the CryoAC geometric rejection efficiency (98% TBC);
- Operation at $T_B$ = 50-55 mK, corresponding to the X-IFU thermal bath;
- Reproducible thermal conductance between the absorber and the thermal bath, in order to control the thermal decay time of the detector, which defines its dead time. At present, this implies for the design a suspended absorber.

### 2.1 Design

The CryoAC DM is based on a wide area silicon absorber (1 cm$^2$, 525 µm thick) sensed by a network of 96 iridium:gold TES in parallel configuration, and readout by a SQUID operating in the standard Flux Locked Loop (FLL) configuration. To obtain a well-defined and reproducible thermal conductance towards the thermal bath, the absorber is connected to a silicon rim (gold plated and in strong thermal contact with the bath) through four narrow silicon bridges (100 x 1000 µm$^2$), achieving a suspended free-standing structure (Fig. 2).

The TES network is designed to ensure a uniform absorber surface coverage for an efficient athermal phonons collection, while constraining down the heat capacity contribution of the metal film thanks to the quite small TES size (50x500 µm$^2$). All these design solutions are necessary to speed up the detector thus enabling the high rejection efficiency of particle background imposed by mission requirements. The bias line on board of the absorber are anti-inductive overlapping Nb wirings, to limit the magnetic coupling with the TES array.

Platinum heaters are also embedded on the absorber to increase its temperature if necessary. This has been foreseen to deal with the high current needed to bias the TES network with $T_B$=50 mK, that has been measured to be of the order of mA in the previous CryoAC prototype [4]. By means of the heater it will be possible to operate the detector by lower bias current, locally increasing the absorber temperature, thus limiting magnetic coupling effects to the TES array. Furthermore, increasing the absorber temperature could be useful to increase the heat capacity of the detector, extending its maximum detectable energy ($E_{MAX} \sim C \cdot \Delta T_C$). Alternatively, the heater can also be switched on to drive the TES into the normal state and bias it, thus avoiding to overcome the high critical current by injecting on the detector several mA, and then switched off during the detector operations once the working point is got.

The specifications of the CryoAC DM are summarized in Tab. 1.

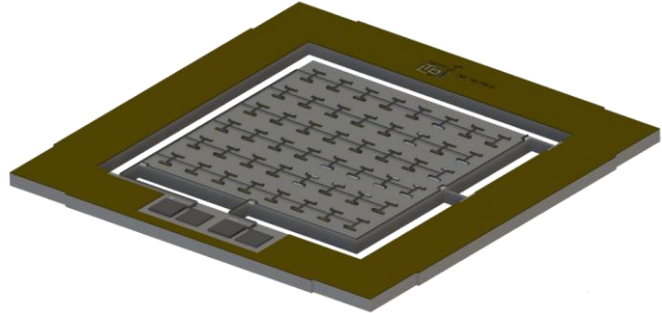

Figure 2. Sketch of the CryoAC DM design.

Table 1. CryoAC DM specifications

| Parameter | Value |
|---|---|
| Silicon chip thickness | 525 μm |
| Silicon chip resistivity | > 10 kΩ·cm |
| Chip overall area | (16.6 × 16.6) mm² |
| Absorber area | (10.0 × 10.0) mm² |
| Beam dimensions | (1000 × 100) μm² |
| Rim width | 2.3 mm |
| Ir/Au TES size (× 96) | 50 × 500 μm² (~ 150 nm thick) |
| Ir/Au TES $T_C$ | 100 mK (TBC) |

## 2.2 Fabrication steps and status

The detector is produced from a commercially available silicon wafer at the Genoa University (Phys. department). The fabrication process is divided as follows:

A. Ir:Au film deposition on silicon chip by Pulsed Laser Deposition;
B. Ir:Au film patterning using positive photolithography and dry etching process (Fig. 3B);
C. Pt heaters deposition by e-beam evaporation, patterning performed using negative photolithography and lift-off process (Fig. 3C);
D. Nb wiring (lower strip) deposition by RF-sputtering, pattering performed using negative photolithography and lift-off process (Fig. 3D);
E. SiO$_2$ insulation layer deposition by e-beam evaporation, patterning performed using negative photolithography and lift-off process (Fig. 3E);
F. Nb Wiring (upper strip) deposition by RF-sputtering, patterning performed using negative photolithography and lift-off process (Fig. 3F);
G. Au deposition on the rim by thermal evaporation, patterning performed using negative photolithography and lift-off process;
H. Si Deep RIE-ICP etching using bosch process using aluminum Hard-mask deposition by thermal evaporation, patterning performed using negative photolithography and lift-off process (Fig. 3H);

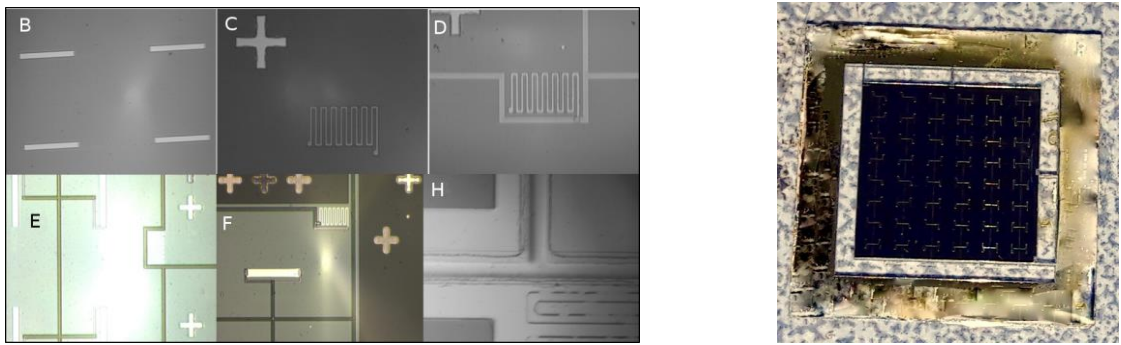

Figure 3. (Left) CryoAC DM fabrication steps. See text for details. (Right) Sample at the end of the fabrication process (not working prototype due to an issue on the bias pads).

Some pictures representing the different steps of the fabrication process are shown in Fig. 3 - Left.

At present, all the processes until final etching (from Step A to Step H) have been defined and tested. During the processes setting, some defective DM sample has been produced (Fig. 3 - Right) and used to define and test the handling and integration procedures. About the last step, the Si RIE-ICP final etching (Step H) turned out to be poisoned by the RIE-ICP of Ir:Au (step B), with decreases of etching rate and quality with respect to our first tests. We are now setting the two etching processes (step B and step H) in two different reactors to avoid cross-contamination, and we have started to etch dummy samples to recover the original RIE-ICP process quality. These dummy samples are then thermal cycled in the dilution unit to test their robustness and the handling/integration procedure.

In the meantime, we have decided to characterize and test a DM-like sample (namely AC-S9) before the final etching, thus seeing if the process will affect the DM. This has allowed us also to verify all the previous fabrication steps and to preliminary test the detector integration and our cryogenic setup.

## 3. AC-S9 & THE DM-LIKE SETUP

The pre-etching AC-S9 DM sample has been integrated and tested in a DM-like setup (Fig. 4).

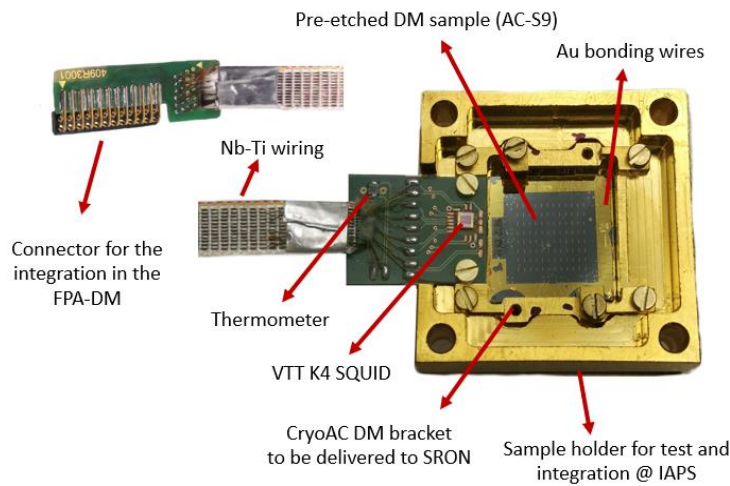

Figure 4. The pre-etching DM sample (AC-S9) integrated and tested in a DM-like setup. The DM CryoAC will appear as this sample but etched.

The proper CryoAC DM assembly consists of the detector sample and a PCB with its cold read out electronics (based on a VTT K4 SQUID having an on-board $R_S$=0.5 m$\Omega$ shunt resistor, and including a SMD Ruthenium Oxide thermometer). Both the detector and the PCB are integrated on a dedicated OFHC copper gold-plated bracket which will be delivered to SRON for the integration in the FPA DM. The CryoAC DM assembly will be placed on the backside of the TES array mount plate, leaving a clearance of 0.5 mm between the TES array and the CryoAC DM detector (Fig. 1). In our test setup the bracket is instead inserted in a dedicated gold plated OFHC copper sample holder (Fig. 4). A Nb-Ti loom runs from the cold electronics to a PCB connector designed by SRON to fit the FPA DM setup. For thermalization purpose, gold bonding wires are applied from the gold-plated detector rim to the supporting bracket.

We remark that for AC-S9, since it has been not etched, these Au wire bondings represent the main contribution to the thermal conductance between the silicon absorber and the thermal bath (i.e. the bracket and the sample holder assembly).

## 4. AC-S9 PRELIMINARY CHARACTERIZATION

The AC-S9 DM-like setup has been integrated and tested in the INAF/IAPS Roma Cryolab, using a dry cryogenic system (Dilution Refrigerator precooled with a Pulse Tube). The system has been magnetically shielded at cold with a Cryophy shield anchored to the Mixing Chamber plate. The detector has been operated with a commercial 2-channels Magnicon XXF-1 electronics, which has been used to operate both the TES/SQUID system (channel 1) and the on-board heater (channel 2 current generator).

The transition curve of the sample is shown in Fig. 5-Left. The narrow transition ($\Delta T_C < 1$ mK), obtained with 96 TES connected in parallel, demonstrates the uniformity of the TES network. The critical temperature ($T_C = 163$ mK) and the normal resistance ($R_N = 40$ m$\Omega$) are instead significantly higher than the expected values ($T_{C, target} \sim 100$ mK, $R_{N, expected} \sim 10$ m$\Omega$), revealing a problem occurred during the TES deposition procedure. A subsequent investigation revealed that this was due to an incorrect calibration of PLD process to deposit the Ir:Au film: the TES are indeed thinner than expected (global thickness ~ 50 nm instead of the ~150 nm goal). As reported in literature [5], the critical temperature of thin Ir film significantly increases with respect to the bulk value ($T_{C, bulk Ir} \sim 140$ mK) for thickness lower than ~100 nm. This has increased our TES $T_C$, even though the proximization of the Ir film, and it has also increased the normal resistance of the network. At present activity is on going to re-calibrate the PLD process by producing and testing some Ir:Au test structures.

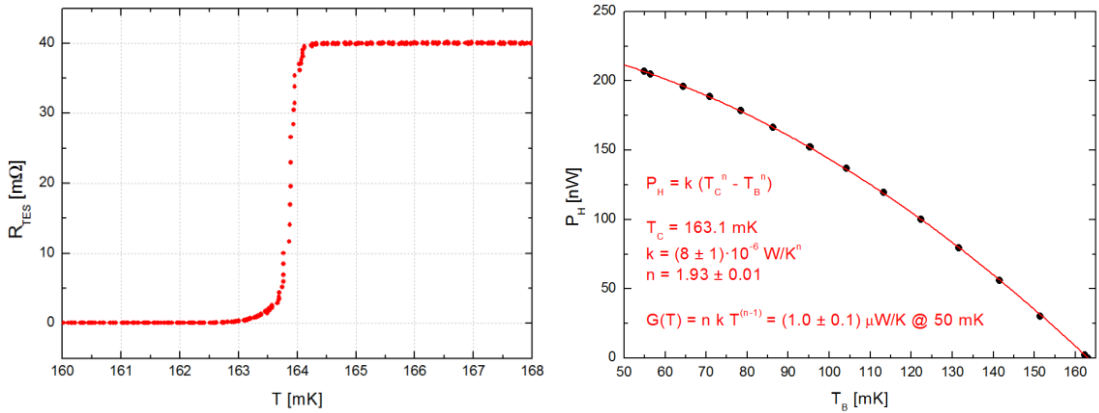

Figure 5. Preliminary thermoelectric characterization of the sample. (Left) Transition curve. (Right) Power injected on the absorber by the on-board heater to bring the TES into the transition ($R_{TES}/R_N = 5\%$) as a function of the thermal bath temperature.

We have then operated the on-board heater to heat the absorber and drive the TES into the transition (at the fixed level $R_{TES}/R_N = 5\%$), at different thermal bath temperatures (Fig. 5 - Right). The experimental points have been fitted with the general equation describing the power flow to the heat bath [6]:

$$P_H = k \cdot (T_C^n - T_B^n) \tag{1}$$

obtaining the best-fit values *n = (1.93 ± 0.01)* and *k = (8 ± 1)·10⁻⁶ W/K*, and a respective thermal conductance:

$$G_{measured} = n \cdot k \cdot T_b^{(n-1)} = (1.0 \pm 0.1)\ \mu W/K\ @\ T_b = 50\ mK \tag{2}$$

As expected, the thermal link between the absorber and the thermal bath is dominated by a metallic component (n ~ 2), corresponding to the gold wire-bondings between the gold plated detector rim and the supporting bracket. It is possible to estimate the thermal conductance of the bonding wires by the formula [7]:

$$G_{Au\ Bondings} = n_B \cdot (v_f \cdot d_e)\ /\ 3 \cdot \Gamma_e \cdot S/L \cdot T = 1.5\ \mu W/K\ @\ T = 50\ mK \tag{3}$$

where we used the parameters listed in Table 2. The measured and the expected values are roughly consistent, validating our description of the thermal conductance between the detector rim and the thermal bath.

To complete the description of the thermal properties of the detector, in Table 3 it is reported the estimation of its heat capacity at 50 mK and 160 mK, evaluated by taking into account the different contributions from TES, absorber and rim gold plating.

Table 2. Gold bonding wires properties

| Parameter | Description | Value |
|---|---|---|
| $n_B$ | Bondings number | 20 |
| $v_{f,\ Au}$ | Fermi velocity | $1.4 \cdot 10^6$ m/s [7] |
| $d_{e,\ Au}$ | Electron diffusion length | 1 μm [7] |
| $\Gamma_{e,\ Au}$ | Electronic heat capacity coefficient | 66 J/(m³·K²) [7] |
| S | Bondings section | $n_B \cdot \pi \cdot (12.5\ \mu m)^2$ |
| L | Bondings length | < 1 cm |

Table 3. AC-S9 heat capacity evaluation (parameters in red are estimated from measurements on a Silicon sample)

| Parameter | Source | Formula | Parameters | Value @50mK | Value @160mK |
|---|---|---|---|---|---|
| $C_{TES,\ el}$ | Electrons in the TES | $2.43 \cdot \rho/A \cdot V \cdot \gamma_e \cdot T$ | $\rho_{Ir} = 22.56$ g/cm³<br>$A_{Ir} = 192.217$ g/mol<br>$\gamma_{e,Ir} = 3.20 \cdot 10^{-3}$ J/(K²·mol) [8]<br>$V = 96 \cdot 500\mu m \cdot 50\mu m \cdot 50nm$ | 6 pJ/K | 19 pJ/K |
| $C_{ABS,\ el}$ | Electrons in the absorber | $\rho/A \cdot V \cdot \gamma_e \cdot T$ | $\rho_{Si} = 2.329$ g/cm³<br>$A_{Si} = 28.085$ g/mol<br>$V = 525\ \mu m \cdot 16.6\ mm \cdot 16.6\ mm$<br>$\gamma_{e,Si} = 5.06 \cdot 10^{-8}$ J/(K²·mol) | 32 pJ/K | 96 pJ/K |
| $C_{ABS,\ ph}$ | Phonons in the absorber | $12/5 \cdot \pi^4 \cdot R \cdot \rho/A \cdot V \cdot T^3/\theta_D^3$ | $R = 8.314472$ J/(mol·K)<br>$\rho_{Si} = 2.329$ g/cm³<br>$A_{Si} = 28.085$ g/mol<br>$V = 525\ \mu m \cdot 16.6\ mm \cdot 16.6\ mm$<br>$\theta_D = 645$ K [9] | 11 pJ/K | 352 pJ/K |
| $C_{GOLD}$ | Rim gold plating | $\Gamma_e \cdot V \cdot T$ | $\Gamma_{e,Au} = 66$ J/(m³·K²) [7]<br>$V = 300$ nm · 127.52 mm² | 126 pJ/K | 403 pJ/K |
| $C_{AC-S9}$ | **Total heat capacity** | $C_{TES,\ el} + C_{ABS,\ el} + C_{ABS,\ ph} + C_{GOLD}$ | | **175 pJ/K** | **870 pJ/K** |

# 5. THERMAL & ATHERMAL RESPONSES

We have stimulated the detector and studied its response in two different ways:

- Injecting energy into the absorber by sending fast square current pulses (width = 1 µs) to the on-board heater (Injected energies: 115 keV, 170 keV, 380 keV, 460 keV, 680 keV, 1000 keV, 1830 keV)
- Illuminating the absorber with X-ray photons by means of collimated radioactive sources (6 keV photons from a $^{55}$Fe source illuminating the absorber surface with the TES network, and 60 keV photons from a shielded $^{241}$Am source)

All the acquisitions have been made fixing $T_B$ = 160 mK and biasing the detector at $R_{TES}/R_N$ ~ 5 % ($I_{TES}$ = 325 µA), with negligible loop gain factor. The SQUID output has been filtered by a low noise band-pass filter 1Hz-10kHz, and acquired with an ADC sampling rate of 500 ks/s.

In Fig. 6 - Left it is reported the shape of the acquired average pulses normalized by their areas. Note that the pulse shape strongly depends from the method to inject energy into the absorber (photon absorption / heater injection), whereas - fixed the method - it is quite independent from the energy amount.

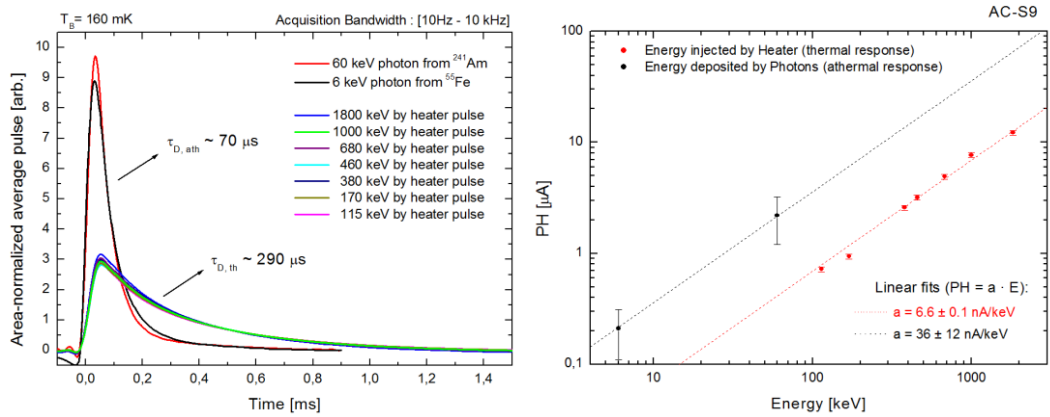

Figure 6. Study of the detector thermal and athermal responses. (Left) Shape of the average pulses acquired stimulating the detector with X-ray photons and fast heater pulses. (Right) Average Pulse Height as a function of the energy injected by the heater (red points) or deposited by the X-ray photons (black points).

The detector response to the heater energy injections is purely thermal, showing a decay time $\tau_{D,TH}$ ~ 290 µs roughly consistent with the expected C/G ratio at 160 mK ( C/G ~ 870 [pJ/K] / 3.2 [µW/K] ~ 270 µs), thus showing the negligible ETF loop gain.

As expected, the response to the photon absorption is instead faster than the thermal one, demonstrating that the detector is able to work in an athermal regime. The TES network is indeed able to collect the first population of phonons generated when a particle or photon deposits energy into the absorber, before their thermalization into the silicon. Note that in this case the pulses do not show any significant thermal component, differently from what we observed in the past CryoAC prototypes [4][10]. We are now investigating this point, and we are carrying on an activity aimed to develop a model able to describe the detector thermal and athermal dynamics.

In Fig. 6 - Right the Average Pulse Height is plotted as a function of the energy, with the error bars showing the width of the PH distributions (1σ). In both the thermal (red points) and athermal (black points) regimes the detector shows a linear response in the explored energy range, as highlighted by the respective linear fits (constrained to pass through the origin). Note that working in the athermal regime allows to increase the detector response, thus lowering the low energy threshold on the detector. On the other hand, the athermal PH dispersion is higher with respect to the thermal one, reducing the spectroscopic capabilities of the detector.

# 6. ENERGY SPECTRA

The energy spectra acquired illuminating the detector with the $^{55}$Fe and the $^{241}$Am sources are shown in Fig. 7. They have been obtained from the raw PH spectra, where the energy axis has been calibrated assuming an athermal detector response of 36 nA/keV (from Fig. 6 - Right). The dashed blue line in the plot represents the trigger threshold (5σ baseline level), whereas the green lines are Gaussian fits of the main spectral bumps.

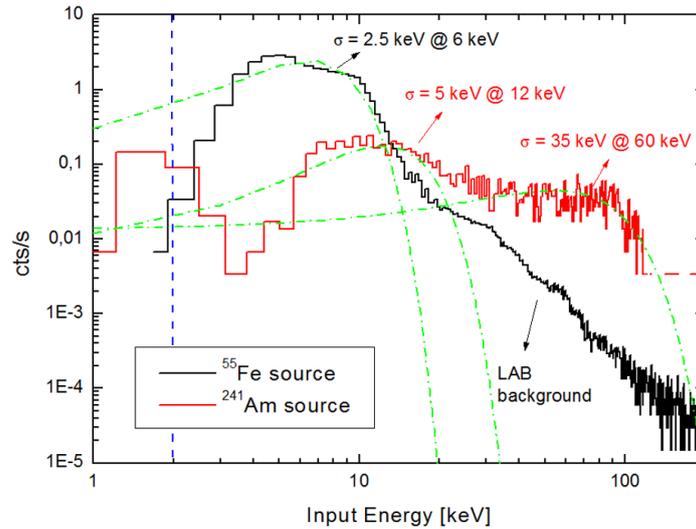

Figure 7. Raw energy spectra acquired illuminating the detector with the $^{55}$Fe (black line) and the $^{241}$Am (red line) sources. The dashed blue line represents the trigger threshold (5σ baseline). The green lines are Gaussian fits of the main spectral bumps (6 keV, 12 keV, 60 keV).

The $^{241}$Am spectrum (red curve) refers to a 5 minutes acquisition, with a count rate of ~ 9 cts/s. It shows two main bumps: the first around 12 keV is due to Compton-scattered 60 keV photons ($E_{Compton\ Edge,\ 60\ keV}$ = 11.3 keV) and probably copper and gold fluorescences ($E_{Cu,k\alpha}$ = 8 keV, $E_{Au,k\alpha}$ = 10 keV). The second one corresponds to the 60 keV photopeak, which is largely spread by the instrumental response.

The "black" spectrum is reconstructed from two different acquisitions. The first one, 5 minutes duration at ~ 36 cts/s and low trigger threshold (~ 2 keV), has been used to detect the main 6 keV line of $^{55}$Fe. The second one has been instead performed with an higher trigger threshold (~ 20 keV) and a much longer acquisition time (~ 19 hours with a count rate ~0.07 cts/s), in order to show the laboratory background level. This was also a stability test of the detector bias, which did not show any significant drift during the long acquisition time.

We remark that being able to detect the $^{55}$Fe 6 keV line, the detector demonstrates a low energy threshold fully compatible with our requirement (20 keV), thus representing a promising step towards the CryoAC DM.

From the spectroscopic point of view, AC-S9 shows instead a very poor energy resolution (ΔE/E ~ 1). This has probably due to the several limitations due to the absence of the final etching (high thermal capacity, high thermal conductance, partial TES surface coverage). From the DM we expect an heat capacity about 5 times lower than the AC-S9 one, an halved thermal conductance and an optimal absorber TES coverage, thus a better detector response. Anyway, we remark that the CryoAC is aimed to operate as anticoincidence particle detector, and it is not conceived to perform X-ray spectroscopy (i.e. there are no requirements on the CryoAC spectral resolution).

## 7. OPERATION AT $T_B$ = 50 mK

Thanks to the onboard heater, we have been able to operate the detector also with the thermal bath at $T_B$ = 50 mK, as required for the DM. We have first fixed the bath temperature, and then used the heater to inject power into the absorber ($P_H$ = 195 nW), increasing its temperature. Finally, we have injected current into the TES circuit until we reached a good working point ($I_B$ = 1000 µA). We can estimate the absorber temperature as:

$$T_{ABS} = [(P_H + P_{TES})/k + T_B^n]^{1/n} \sim 155 \text{ mK} \qquad (4)$$

where $P_{TES}$ is negligible and the k and n parameters are those evaluated in Par. 4.

For comparison we have also performed an acquisition setting directly $T_B$ = 155 mK and biasing the TES with the same bias current ($I_B$ = 1000 µA), without operating the heater. Two signal strips acquired in these different conditions are shown in Fig. 8 - Left. In both the cases, the absorber has been stimulated with the $^{55}$Fe source illuminating the surface with the TES network (see Fig. 9, TES side illumination). The 6 keV pulses indicated by the red arrows are zoomed on the Right of the Figure.

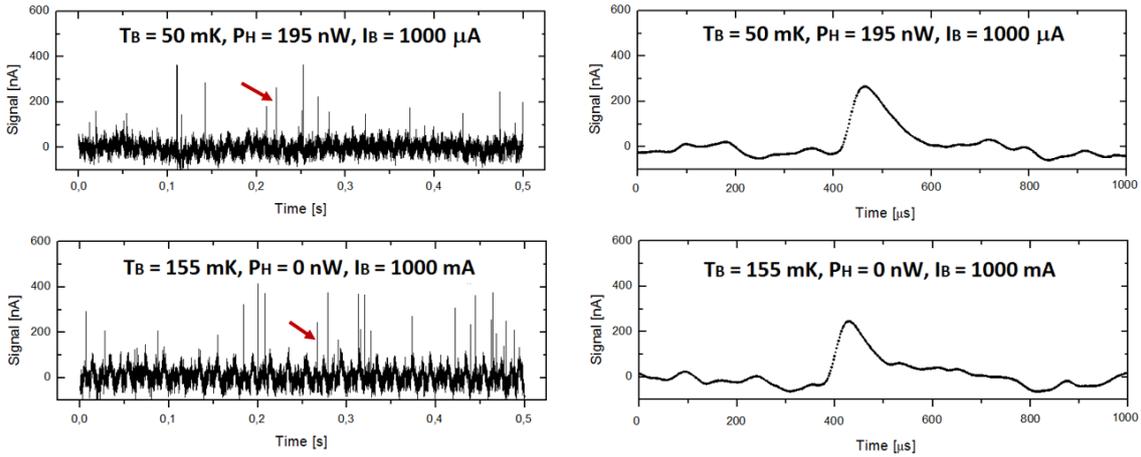

Figure 8. (Left) Signals acquired with the detector working in different conditions, with (top) and without (down) operate the onboard heater. In both the cases the absorber has been stimulated with the $^{55}$Fe source. See text for details. (Right) Zoom on the 6 keV pulses indicated on the left by the red arrows.

The AC-S9 response results very similar in the two different operation conditions, demonstrating that the detector can be operated at $T_B$ = 50 mK without noticeable issues. This is another promising step towards the CryoAC DM.

## 8. $^{55}$FE FRONT/REAR SIDE ILLUMINATION

Lastly, we have stimulated the detector by means of the $^{55}$Fe source alternatively illuminating the TES and the "rear" side of the absorber (see the sketch inside Fig. 9). We remark that 6 keV photons are completely absorbed in ~ 100 µm of silicon (transmission < 5%), so we can expect different paths for the phonons propagation towards the TES in the two cases, thus a different detector response. The Pulse Height spectra acquired with the detector operated in the same working point ($T_B$ = 50 mK, $P_H$ = 195 nW, $I_B$ = 1000 µA) are shown in Fig. 9.

Note that the spectrum is better shaped with the "rear" side illumination, whereas illuminating the absorber from the TES side the Pulse Height distribution shows a long tail after the main peak.

We are now carrying on an activity aimed to simulate the phonons propagation in the absorber and investigate the generation mechanism of the thermal and athermal pulses. This is a due activity in the context of the usual Phase A mission program (trade-off studies).

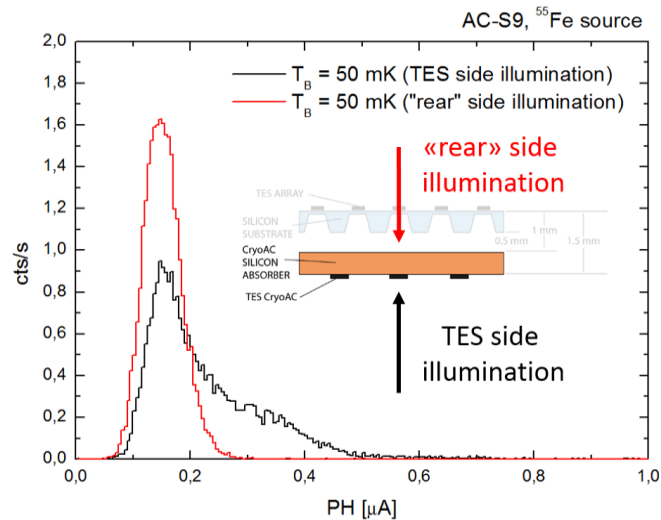

Figure 9. Pulse Height spectra acquired stimulating the detector by means of a $^{55}$Fe source (6 keV photons), illuminating the absorber from the TES (red line) and the "rear" (black line) side.

## 9. CONCLUSIONS

The activity to develop the CryoAC DM is on going. Our goal is to produce a detector compliant with the specifications reported in Table 1, and to deliver it to SRON for its integration in the FPA DM by the end of this year (2018).

The activity performed on AC-S9, a pre-etching CryoAC DM sample integrated and tested in a DM-like setup, has been reported. The results of this activity show promising steps towards the DM, in particular:

- We have successfully operated the heater on-board the detector absorber;
- We have measured the thermal conductance between the DM rim and the thermal bath (due to the Au wire bondings), validating our description of this thermal coupling;
- We have observed and studied both the thermal and athermal detector responses, producing the respective pulse templates.
- Despite the limitations due to the absence of the final etching (high thermal capacity, high thermal conductance, partial TES surface coverage) we have detected 6 keV photons, thus having a low energy threshold fully compatible with our requirement (20 keV);
- We have been able to operate the detector at $T_B$ = 50 mK (DM requirement);
- Illuminating by means of a $^{55}$Fe source the absorber from the rear side (the opposite side with respect to the TES network), a good spectrum shape has been obtained.

## ACKNOWLEDGEMENT

This work has been partially supported by ASI (Italian Space Agency) through the Contract no. 2015-046-R.0, and by ESA (European Space Agency) through the Contract no. 4000114932/15/NL/BW. The research leading to these results has also received funding from the European Union's Horizon 2020 Programme under the AHEAD project (grant agreement n. 654215). The Authors would like to thank the SRON FPA development team for the support in the DM-like setup design and integration.